\documentclass[12pt]{article}
\usepackage{latexsym}
\usepackage{graphicx}
\usepackage{epsfig}
\usepackage{epstopdf}
\usepackage{amsmath}
\usepackage{cite}
\usepackage{hyperref}
\usepackage{amssymb}

\begin{document}

\begin{center}
\textbf{\Large Multivaluedness Aspects in Self-organization, Complexity and Computations Investigations by Strong Anticipation}\footnote{This paper is published in: Kyamakya K. et al. (eds.) \textit{Recent Advances in Nonlinear Dynamics and Synchronization}, Studies in Systems, Decision and Control 109, Springer, Cham, 2018, https://doi.org/10.1007/978-3-319-58996-1$\_$3  }
 \vspace{0.5 cm}

 Alexander Makarenko\footnote{e-mail:
\url{makalex51@gmail.com}} \vspace{0.5 cm}

Institute for Applied System Analysis at National Tech. University of Ukraine (KPI), Kyiv, Ukraine, 
\end{center}

\begin{quote} \textbf{Abstract. }{\small  
Since the introduction of strong anticipation by D.~Dubois the numerous investigations of concrete systems have been proposed. In proposed paper the new examples of discrete dynamical systems with anticipation are considered. The mathematical formulation of problems, possible analytical formulas for solutions and numerical examples of presumable solutions are proposed.
One of the most interesting properties in such systems is presumable multivaluedness of the solutions. It can be considered from the point of view of dynamical chaos and complex behavior. We represent examples of periodic and complex solutions, attractor's properties and presumable applications in self-organization.
The main peculiarity is the strong anticipation property. General new possibilities are the presumable multivaluedness of the dynamics of automata. Possible interpretations of such behavior of cellular automata are discussed. Further prospects for development of automata theory and hyper computation are proposed.
}
\end{quote}

\begin{quote} \textbf{Keyword: }{\small 
nature-inspired; strong anticipation; multivalued solutions, chaos, self-organization; hypercomputation
}
\end{quote}

\vspace{0.5 cm}

\section{Introduction. Short outlook of single- and multi- valuedness in nature and in models}
\label{sec:1}
Recent science of self-organization is now one of the important and developing branches of nature, living systems, and society investigations. Here we outline as the examples only some achievements: dissipative structures theories and synergetic (I.~Prigogine, G.~Nicolis, H.~Haken), sociodynamics (W.~Weidlich, G.~Haag, D.~Helbing), cellular automata (B.~Chopard, M.~Droz, S.~Wolfram, L.~Chua), morphogenesis theory (A.~Turing), artificial neuronets (J.~Hopfield) and many others.

Now a lot of models of such objects and their solution exist: parabolic equations of heat and mass transfer; kinetic equations; equations with memory and space non-locality; ordinary differential equations; cellular automata; discrete equations etc. Such models have a variety of solutions: stationary and periodic solutions, deterministic chaos, solitons and autowaves, `chimera' solutions, blow-up (collapses), synchronization, fragmentation, `ideal' turbulence by A.~Sharkovski and colleagues (see for examples, P.~Shuster, A.~Pikovski, Yu.~Maystrenko, E.~Mosekilde, A.~Sharkovski, A.~Samarski \& S.~Kurdiumov, P.~Sloot, A.~Hoekstra). In most of the cases, since I.~Newton, one of the main requirements is the \emph{single-valuedness} of the solutions in suitable spaces of the solutions (frequently in very complicated spaces, for example in Sobolev's spaces).

At the same time the understanding of computational processes in nature and artificial devices have been received. The examples are automata theory (M.~Sipser, B.~Cooper), cellular automata theory (J.~von Neumann, S.~Wolfram, L.~Chua, A.~Illiachinski). The applications are distributed from quantum level (G.~t~'Hooft, A.~Zellinger, G.~Grossing, H.-T.~Elze, K.~Zuse, S.~Wolfram). Remark that before most of such objects of computational theory and applications also were single-valued (classical automata theory, computer architecture, etc.).

However in parallel to the problems with single-valued solutions many phenomena were found in nature and engineering, which have models with \emph{multi-valued solutions} (that is the existence of presumable many values of solution at given moment of time). Remark that also the mathematical tools for considering multi-valued solutions have been developed. As the examples of tools we can recall variational inequalities, differential inclusions, differential equations with discontinuous coefficients and nonlinear sources. The number of the models with intrinsically multivalued solution have been proposed recently. The examples are multivalued fields (H.~Kleinert), mechanical systems with discontinuities (M.~Zak, A.~Ioffe), turbulence (J.~Leray), multi-valued solitons (V.~Vachnenko), control theory and differential games (A.~Chikriy, R.~T.~Rockafellar), non-classical mechanical systems with friction and collisions (J.~Moreau, C.~Glocker), systems with multi-valued Hamiltonians  in the fields theory (M.~Henneaux, C.~Teitelboim, J.~Zanelli), multivalued functionals in the fields theory (S.~Novikov), non-uniqueness phenomena (ghost) in the fields theory (L.~Faddeev, A.~Grib), Hamiltonian inclusions system (R.~T.~Rockafellar) and of course inverse problems.

But recently it has been found that the investigation of the systems with anticipation is also very interesting. The term `anticipation' was firstly attached to the systems with the intrinsic models for predicting the evolution of the systems and of their environment (see \cite{Rosen}, \cite{Dubois1997, Dubois2001}, \cite{LazarenkoMakarenko}-\cite{Makarenko2011}, \cite{MakarenkoStashenko}-\cite{ScharkovskiEtc} and many others). But now since the works by Daniel M.~Dubois (Belgium) the notion of `strong anticipation' has been introduced and investigated. In case of strong anticipation the system don't have the models for predictions but are self-making with accounting presumable future states of the system. D.~Dubois also introduced the notion of incursive and hyperincursive systems (with presumable multivaluedness of solutions). Also D.~Dubois firstly described the elementary single element with hyperincursion. One of the most important examples of the models with anticipation follows from modeling of large social systems. Some examples of investigations of the systems with strong anticipation were described in \cite{LazarenkoMakarenko}-\cite{Makarenko2011}, \cite{MakarenkoStashenko}-\cite{ScharkovskiEtc}. Remark that one of the most new and interesting properties in such systems is presumable multivaluednes of the solutions. Because of this it is very prospective to consider the properties of multivalued solutions from the point of view of synchronization investigations.

Thus the multivaluedness as the phenomena has many manifestations at least at mathematical models of different natural and artificial objects. So further investigations of presumable multivaluednesss in mathematical problems are very important and interesting. Moreover the manifestation and interpretations of multivaluedness can be important for further understanding of self-organization, computation theory, living system properties including consciousness, observability and measurements, complexity and others. In fact presumably all the systems and problems which have been considered previously on the base of models with single-valued solutions can have the counterpart with multivalued solutions. Especially interesting can be interpretations related to the nature and society behavior.

Systems with anticipation (with advanced effects) constitute one of the classes of the systems with presumable multivaluedness. Remark that just now many examples of the systems with anticipation exist: electromagnetic theory (R.~Feynman, D.~Dubois), nonlocal fields theory (N.~Nielsen), economic problems (L.~Gardini), control problems, evolution on the lattices, neuron models, social systems (A.~Makarenko, L.~Leydersdorff \& D.~Dubois), neuronets and consciousness (A.~Ma\-ka\-renko) and many others. But till now the systems with anticipation (especially with strong anticipation) haven't been considered from the point of view of relevancy to the self-organization theory and computation theory. So the first goals of given paper is to attract attention to such problems. For this task we describe some examples of multivalued behavior in the systems with strong anticipation. Also we propose the discussion on presumable development of investigation and possible achievements.

The structure of the paper is the next. In Section~\ref{sec:2} we give the short description of strong anticipation. Section~\ref{sec:3} is devoted to the short review of different systems with strong anticipation: cellular automata, neural networks, discrete equations, partial differential equations and to the illustration of their multivalued solutions. Discussion of multivalued solution properties and presumable research problems is presented in Section~\ref{sec:4}. Sections~\ref{sec:5}-\ref{sec:7} deal with some possible consequences of strong anticipation in computation theory, consciousness theory and uncertainty problem. Because of the absence of the space we pose only the short description of examples and ideas. Thus the main goal of this paper is to give the general outlook on the presumable multivaluedness in models with strong anticipation, their interpretations and further prospects for investigations.

\section{Strong anticipation property}
\label{sec:2}
Since the beginning of 90-th in the works by D.~Dubois -- see \cite{Dubois1997, Dubois2001} the idea of strong anticipation had been introduced: `Definition of an incursive discrete strong anticipatory system \ldots: an incursive discrete system is a system which computes its current state at time $t$, as a function of its states at past times $\ldots,t-3,t-2,t-1$, present time  $t$, and even its states at future times $t+1,t+2,t+3,\ldots$
\begin{equation}
\label{eq:1}
x(t+1)=A(\ldots ,x(t-2),x(t-1),x(t),x(t+1),x(t+2), \ldots ,p),
\end{equation}
where the variable $x$ at future times $t+1,t+2,t+3, \ldots$ is computed in using the equation itself.

Definition of an incursive discrete weak anticipatory system: an incursive discrete system is a system which computes its current state at time $t$, as a function of its states at past times  $\ldots,t-3,t-2,t-1$, present time $t$, and even its predicted states at future times $t+1,t+2,t+3,\ldots$
\begin{equation}
\label{eq:2}
x(t+1)=A(\ldots ,x(t-2),x(t-1),x(t),x^*(t+1),x^*(t+2), \ldots ,p),
\end{equation}
where the variable $x^*$ at future times $t+1,t+2,t+3,\ldots$ are computed in using the predictive model of the system' \cite{Dubois1997}, (D.~Dubois).

Many results on the systems with strong anticipation have been described in the papers by D.~Dubois, his co-authors and followers (see \cite{CASYS}). Such investigations originated from specific for computer science formulations. But the great variety of such systems opens the new possibilities for investigations. One of them is self-organization or synergetics. So in this paper we will try to point out such possibilities especially with presumable multivaluedness of solutions.

Thus as the further research problem in the field of self-organization should be considered in the system with strong anticipation. Remark that the simplest but rather general counterparts for Eq.~\ref{eq:1} and Eq.~\ref{eq:2} are the equations for continuous time case.

\section{Examples of investigated problems with anticipation}
\label{sec:3}

According to the general goal of the paper in this section we give the collection of different examples of strong anticipatory systems to illustrate the emerging of new peculiarities of the solutions. Here we prefer to give only some illustrations because of lack of space.

In this section we briefly describe for illustration the possibilities of strong anticipation property manifestation in some problems.

\subsection{Two-Steps Discrete-Time Anticipatory Models}
\label{subsec:31}

The proposed model has two features (see details at \cite{MakarenkoStashenko}). The first is two-steps nature (that is passing on two steps ahead at discrete). The second is that the function $f(x)$ in the model has a piecewise-linear character, and looks like the transition function of neurons in neuronets \cite{Makarenko2011}. Remark that the piecewise character of the nonlinearity usually allows to simplify mathematical investigations.

We write down the proposed model in discrete time-step as follows (here $p_{t+1}$ is the value of the variable $p$ at $(t+1)$ moment of time $t=0,1,2, \ldots$)
\begin{eqnarray*}
p_{t+1} &=& \alpha \cdot m+(1- \alpha)p_t - \alpha \cdot f(p_{t+2}-p_{t+1}),
\\
p_{t+2} &=& \alpha \cdot m+(1- \alpha)p_{t+1} - \alpha \cdot f(p_{t+2}-p_{t+1}).
\end{eqnarray*}

The function $f(x)$ depends on the parameter $\alpha$ and has the following expression:
\begin{eqnarray*}
f(x)&=&0 \quad\quad\; x \leq 0,
\\
f(x)&=& \alpha \cdot x \quad x \in \left(0, \frac{1}{\alpha}\right],
\\
f(x)&=&1 \quad\quad\; x> \frac{1}{\alpha}.
\end{eqnarray*}

\begin{figure}[h]
\centering
\includegraphics[width=300pt
]{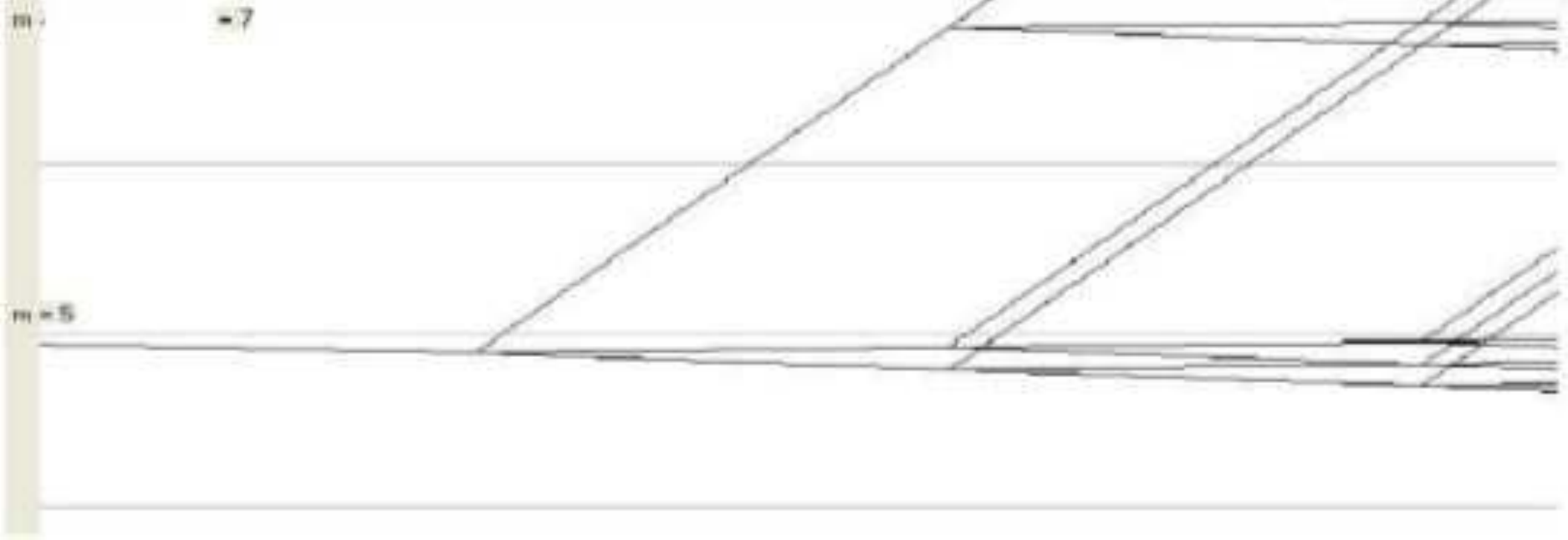}
\caption{The graph of branching of solution. On the $x$-axis we display the discrete time steps and at the $y$-axe we display the $p_{t}$ values.The parameters values are: $\alpha =2$, $m=0$. Limitations for visualization of the values of variables are the next: maximal $m+7$ and minimal $m-5$, without representing the branches, which are thrown out to infinity.}
\label{fig:1}
\end{figure}

\begin{figure}[h]
\centering
\includegraphics[width=130pt]{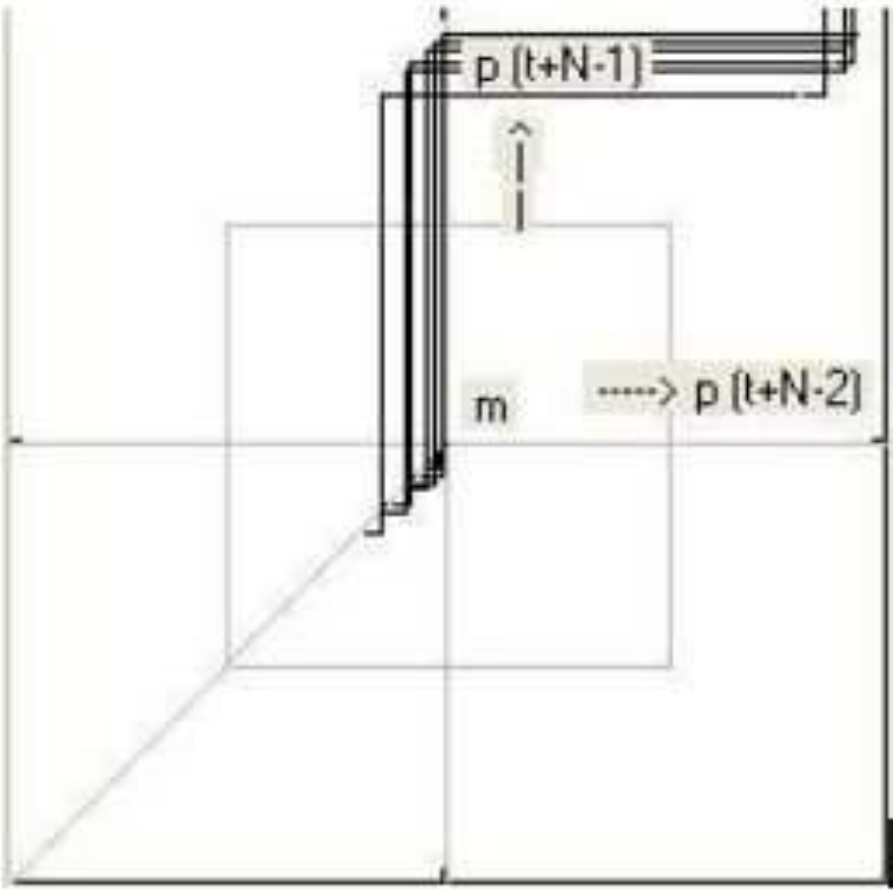}
\caption{Map in state space. The sequence of $(p_t,p_{t+1})$ values on the plane is represented for the solution, which corresponds to the parameters and initial conditions from Fig.~\ref{fig:1}.}
\label{fig:2}
\end{figure}

The software has been developed so that it is possible to visualize some branches of the states of map, but only those, that are not thrown out to infinity (see \cite{MakarenkoStashenko}). This facilitates the understanding of the processes, which take place in the region of ambiguousness. In case of equations given above we find three-folding of solution (two branches tend to infinity and one branch stays in restricted region of space). At Fig.~\ref{fig:1} we display the branching of solution at discrete time steps. We display the discrete time steps on the $x$-axis and the value of $p$ on the $y$-axis. In other calculations we have seen the increasing periods of cycles and increasing of the number of different branches in the course of time, that is we have seen a tendency to phenomena which we can call `chaos' (see Fig.~\ref{fig:2}).

At Fig.~\ref{fig:2} we can see the similar behavior of solutions on different branches.

\subsection{Logistic equation with strong anticipation \cite{LazarenkoMakarenko}}
\label{subsec:32}

Other well-known example of discrete time equations with complex behavior is classical logistic  (or Ferhulst) equation (see for example \cite{ScharkovskiEtc}). It was very interesting to consider the counterparts to this equation with strong anticipation accounting.

The discrete dynamics equation with strong anticipation which is the modification of well-known logistic equations was considered. The proposed equation had the form:
\begin{equation}
\label{eq:3}
x_{n+1} = \lambda x_n (1-x_n)- \alpha x_{n+1}^2,
\end{equation}
where $\alpha \neq 0$ (for $\alpha =0$  we have a classical logistic map) is an anticipatory factor. Multivalued behavior and fractals were found. The examples of multivalued periodic solutions and chaos were considered. Also multivalued fractal properties and bifurcations were investigated.

\subsection{Cellular automata with anticipation}
\label{subsec:33}

Other very general class of models is cellular automata. The basic issues of classical cellular automata approach are: regular subdivision of space on equal (regular) cells; discrete time for considering the evolution of cell states and special (local) rules for dynamics of cell states (see \cite{Wolfram}, \cite{Ilachinski}, \cite{MakarenkoGoldengorinKrushinsky}). Recently we have investigated some kind of such models with anticipation accounting: namely game `Life' with anticipation; movement of pedestrian crowds; general problems \cite{MakarenkoGoldengorinKrushinsky}.

First of all let's recall a short definition of cellular automata (CA). The cellular automata is discrete dynamic system which represents the set of identical cells and connections between them. Cells create a lattice of the cellular automata. Lattices can be different types, differing both on the dimension, and under the form of cells~\cite{MakarenkoGoldengorinKrushinsky}.

The local rule for cell $k$ on the $Z_d$ is the transformation $T_k$ which transforms the state $s_k(t)$ in $S$ (here $S$ denotes the space of all states of the cell) of cell with index $k$ at moment $t$ to the state $s_k(t+1)$ in S of the same cell at moment $(t+1)$. Namely,
\begin{equation}
\label{eq:4}
s_i(t+1)=G_i(\{ s_i(t) \} ,R),
\end{equation}
where $N_k$ is some neighborhood of cell $k$ on the lattice $Z_d$; $\{ s_k(t) \}$ is the set of cell's states within $N_k$, the transformation $T_k$ result depends only on the states of elements within the neighborhood $N_k$ (locality), $R$ -- some parameters for rules that define transformations. The collection of states of cells at given time moment is called 'configuration' (we note it as $C(t)$). The collection of local transformations $T_k$ defines the global transformation $G$ on the configuration space $C$, where $C$ is the space of all cells states collection: $C(t+1)=G(C(t))$.

The initial data $C(0)$ configuration is defined at initial moment $t=0$. The set of transformations $\{ T_k \}$ or transformation $G$ defines the cellular automata on the lattice $Z_d$ with the cell's state space $S$.

In case of strong anticipation accounting Eq.~\ref{eq:4} in additive case can have the form:
\begin{equation}
\label{eq:4a}
s_i(t+1)=(1-\alpha) G_i(\{ s_i(t) \} ,R) + \alpha G_i(\{ s_i(t+1) \} ,R).
\end{equation}
The factor $\alpha$ corresponds to the accounting of strong anticipation. The case $\alpha=0$ corresponds to the absence of anticipation.

\subsection{Game `Life' as example of CA}
\label{subsec:34}

One of the basic examples of CA is well-known Game `Life' by J.~Conway. At a time $t$ let some subset of the cells in the array be alive. In such case $S=\{ 0,1 \}$, where `1' corresponds to the `living' state and `0' to the `dead' one in some interpretations. The living cells at the time $t+1$ are determined by the ones at time $t$ according to the following evolutionary rules. Remark that strong anticipation follows to the origin of equations of type:
\begin{equation}
\label{eq:5}
s_i(t+1)=G_i(\{ s_i(t) \}, \ldots ,\{ s_i(t+g(i)) \},R).
\end{equation}

The main peculiarities of such equation is that such equations can have multivalued solutions (of course in some range of parameters).

\begin{figure}[ht]
\centering
\includegraphics[width=200pt]{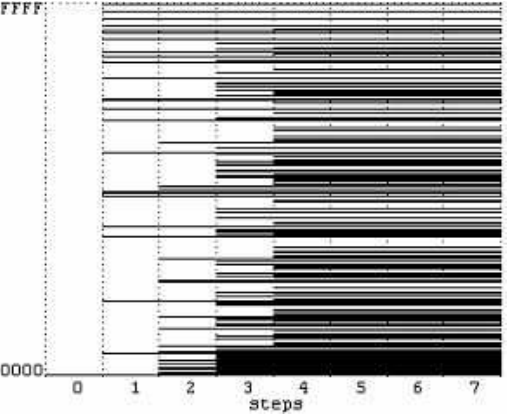}
\caption{A large number of configurations coexisting in system (additive next state function, $\alpha =0.5$, initial state is $0000$). Discrete time moments are represented on the $x$-axis. The configuration's index of $16$ cells collection is represented on the $y$-axis in special alphabet.}
\label{fig:3}
\end{figure}

Fig.~\ref{fig:3} presents an example of well-developed solution multivaluedness within the model `LifeA'. On the $x$-axis we display the values of discrete time steps. On the $y$-axis we show the configurations, which exist at each given moment of time. The configurations are marked by special indexes from $0000$ to FFFF in special alphabet. Fig.~\ref{fig:3} corresponds to the cellular automata with $16$ cells and periodic boundary conditions ~\cite{MakarenkoGoldengorinKrushinsky}.

Such behavior is very interesting and prospective for investigations of self-organization. Remark that the main peculiarities in such investigated systems also are presumable multivaluedness and space non-homogeneity of solutions behavior.

\subsection{Neural networks with anticipation}
\label{subsec:35}

\begin{figure}[ht]
\centering
\includegraphics[width=210pt]{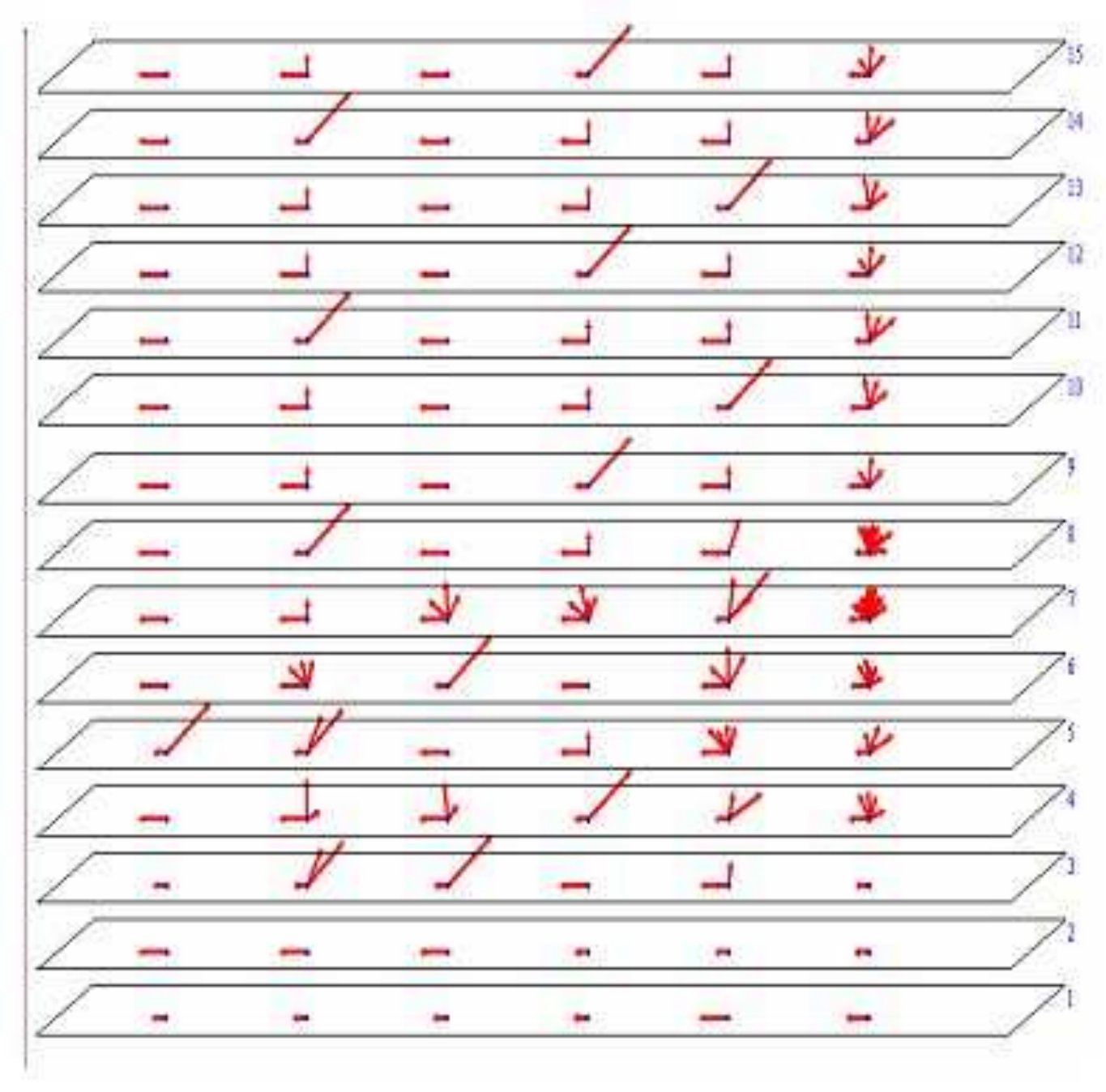}
\caption{Behavior of multivalued solution. The numbers at right side of the figure correspond to the discrete time moments. The network had $6$ elements. The non-homogeneous in space solution is seen. The length of an arrow corresponds to the value of neuron's state. Many arrows origing from given point at the figure correspond to multiple values of cell's state. The length of the arrow corresponds to the value of solution on given branch at given time moment.}
\label{fig:4}
\end{figure}

Other very important object of proposed type (with anticipation and discrete time dynamic) is neural networks with strong anticipation. We have already investigated some such models \cite{Makarenko2011}. Here we propose only very short description of calculations for understanding of presumable behavior. One of the simplest variant of such models with strong anticipation accounting has the form (counterpart for Hopfield's neural networks)
\begin{equation}
\label{eq:6}
x_j(n+1)=f \left( (1- \alpha) \sum \limits_{i=1}^N w_{ji} x_i(n)+ \alpha \sum \limits_{i=1}^N w_{ji} x_i(n+1) \right) .
\end{equation}
Here $x_j(n)$ is the value of $j$-th element at the moment $n$, $w_{ji}$ is the weight matrix and $\alpha$ is anticipation parameter. The case of $\alpha =0$ corresponds to the absence of anticipation (with classical neuronet). At Fig.~\ref{fig:4} we propose the example of presumable behavior of the Eq.~\ref{eq:6} model at different time moments.

Two properties of artificial neural network solutions are represented at Fig. \ref{fig:4}: multivaluedness and non-homogeneities of elements behavior. Both are rather new and prospective for development of synchronization investigations both in theory and in applications.

\subsection{Boundary-value problems with anticipatory boundary conditions}
\label{subsec:36}

Distributed systems with continuous space and time variables constitute the special field of investigation in self-organization theory (since the works of I.~Prigogine, H.~Haken, A.~Kolmogorov, A.~Turing). The investigation of strong anticipation accounting in such systems is only on the initial stage. So in this subsection and in Section~\ref{sec:4} we pose the description of some research problems and interpretations of solutions in multivalued case for distributed systems.

The new examples of distributed systems with strong anticipation have been considered, namely the system of hyperbolic differential equations with special boundary conditions that include the strong anticipation~\cite{Makarenko2010}. The mathematical formulation of problems, analytical formulas for solutions and interpretations of derived distributed solutions are proposed.

We consider some hyperbolic systems with anticipation. Such problems in classical case originated in considering oscillations of elastic string in bounded domain; oscillations of voltage and flows in transmission lines (see examples in \cite{ScharkovskiEtc}, E.Yu.~Romanenko \& A.~Sharkovski (1999) and references there).

The basic classical problems have the form as follows. We consider the system of two first-order hyperbolic equations in bounded space domain
\begin{eqnarray}
u_t+u_x &=& 0, \label{eq:61}
\\
v_t-v_x &=& 0, \label{eq:62}
\\
(x,t) \in \Omega &=& [0,1] \times (- \infty,+ \infty). \nonumber
\end{eqnarray}
The initial conditions should be added to make the problem solution unique. For simplicity in this paper we add the next ones:
\begin{eqnarray}
v(0,t)= \phi _1(x), \quad x \in [0,1], \label{eq:63}
\\
u(0,t)= \phi _2(x), \quad x \in [0,1]. \label{eq:64}
\end{eqnarray}
Also the boundary conditions should be considered for $x=0$ and $x=1$ and all $t \ge 0$ for completeness. Very interesting results were obtained with the following boundary conditions in the works of A.~Scharkovski with colleagues \cite{ScharkovskiEtc}:
\begin{eqnarray}
u(0,t)=v(0,t), \quad t \in (- \infty,+ \infty), \label{eq:65}
\\
v(1,t)=f(u(1,t)), \quad t \in (- \infty,+ \infty). \label{eq:66}
\end{eqnarray}
To discuss the presumable effects of strong anticipation we restrict the consideration in this paper to the following form of boundary condition at the right point of the segment $x=1$
\begin{equation}
\label{eq:67}
v(1,t+1)=f(u(1,t+1),u(0,t+2)).
\end{equation}
Eq.~\ref{eq:67} replaces the condition Eq.~\ref{eq:66}. Thus we need to study the equation
\begin{equation}
\label{eq:68}
u(0,t+2)=f(u(0,t),u(0,t+2))
\end{equation}
similarly as it was done for the problem Eq.~\ref{eq:61} -- Eq.~\ref{eq:66}.

Eq.~\ref{eq:68} can be a very complex object. Due to presumable multivaluedness of solution the strict mathematical formulation of results will be forthcoming.

We propose multivalued counterpart for complex behavior of solutions and their `ideal' turbulence from \cite{ScharkovskiEtc}. The main new property is the possibility of multivalued fields origing (that is the existence of solutions of partial differential equations with many values at each space point at given time moment). This can be interesting for fields equations, for example Dirac equations.

Since \cite{ScharkovskiEtc} it is known that the classical system without anticipation can have very interesting and complex behavior: periodic, relaxation oscillations, oscillations with decreasing characteristic length and chaotic regimes ('ideal' turbulence). But for examples with strong anticipation we can observe different behavior on different branches, multivalued chaotical solutions etc.

\section{Presumable properties of solutions and further research problems in physics}
\label{sec:4}

\subsection{Multivaluedness and self-organization}
\label{subsec:41}

Some generalizations for cases with accounting of strong anticipating can be proposed using the results described in Section~\ref{sec:3}. At first the solutions in such cases can be multivalued as we described above. So the fractal dimension of solutions can be greater than the one in single-valued case. Moreover the behavior of the solutions can be much more complex. Different regimes on the different branches of solutions can exist (some examples of such different branches see in \cite{MakarenkoStashenko}). So each branch can have different stochastic properties, different types of relaxation behavior etc. But propably the most interesting is the common behavior of mixture of the branches in multivalued case. Especially interesting is the problem of limiting behavior as $t \to \infty$: limit solutions, attractors of such solutions, probability properties of limiting objects. But the new problems also arise: observability of solutions, selection the single-valued trajectories of the system, its limiting objects complexity and such complexity measures, searching adequate mathematical spaces for considering the problems and solutions. In particular, the problems of adequate definition of periodic behavior (and moreover of dynamical chaos), Lemeray's staircase, Poincare's bifurcation diagram, ergodicity, mixing are interesting. Also many other problems from self-organization theory \cite{DanilenkoEtc} can be re-considered with strong anticipation.

\subsection{General distributed media with anticipation}
\label{subsec:42}

Much more models can be proposed for the case of considering the continuous media with some kind of anticipation. The simplest variant is when only the nonlinear source $f(u)$ in equations of such media (where $u(x,t)$ is field value) has anticipatory property, that is
\begin{equation}
f_1=f(u(x,t),u(x,t+ \tau ))
\end{equation}
in simplest case or
\begin{equation}
f_2= \int \limits_{- \infty}^{+ \infty} f(u(x, \tau))K(x,t- \tau )d \tau
\end{equation}
in more complex case. The example of such kind of model is parabolic equation of diffusion with such nonlinear source. As the example of such system we can propose the models with continuous variables of neural activation fields (counterparts for known models of S.~Amari or H.~Wilson \& J.~Cowan). Remark that much more complex and developed models of such kind phenomena are the counterparts to well-known strict models of the media with space non-locality and memory in theoretical physics.

\subsection{Oscillators and chimeras and in media with memory, nonlocality and anticipation. New research possibilities}
\label{subsec:43}

Many new interesting solutions have been found before for distributed media: dynamical chaos, oscillations, autowaves, synchronization and quite recently `chimera' states. Till now mostly the classical parabolic equations of hydrodynamics and diffusion have been used (for example Navier-Stokes equations).

But now it is recognized that more accurate equations with memory and space nonlocality effects should be used (see for example \cite{DanilenkoEtc}). Here we describe some possibilities for posing new research problems. First of all we consider the models with memory (relaxation in hydrodynamics). In such case one of the classes of models is the infinite systems of ordinary differential equations of second order in time. Such systems recall the systems of coupled oscillators. So the problems of energy transitions between different scales receive new solutions (transfer from large to small scales). Secondly the new possibilities supply the accounting of nonlocality, This follows to presumable origin of new `chimera' states in hydrodynamics. And finally accounting of anticipation follows to the possibilities of multivalued `chimera' states in distributed systems.

\subsection{`Chimera' states in the systems with strong anticipation}
\label{subsec:44}

Recall that `chimera' states are highly non-homogeneous transitive solutions when different types of behavior coexist in space \cite{AbramsStrogatz}. The `chimera' states are the solutions in chains of elements or in the distributed media that have the different behavior in different domains of the space. For example such systems can have coexisting domains with chaotic and `smooth' behavior in different places of the space. One of the discrete systems with presumable `chimeras' is the given below \cite{AbramsStrogatz}:
\begin{equation}
\label{eq:7}
z_i^{t+1}=f(z_i^t)+ \frac{\sigma}{2P} \sum \limits_{j=i-P}^{i+P} (f(z_j^t)-f(z_i^t)).
\end{equation}
Here $z_i^t$ is the value of $i$-th element at discrete time, $\sigma$ is a parameter, $f$ is a nonlinear function.

The next example is the chains of oscillators:
\begin{equation}
\label{eq:8}
\frac{d \psi _k(t)}{dt} = \omega - \frac{1}{2R} \sum \limits_{j=k-R}^{k+R} \sin (\psi _k(t)- \psi _j(t)+ \gamma ),
\end{equation}
where $\psi _k(t)$ is the value of $k$-th element at the time moment $t$.

The counterpart example to Eq.~\ref{eq:7} with `chimera' state in case of strong anticipation accounting is
\begin{eqnarray}
z_i^{t+1}=(1- \alpha ) \left( f(z_i^t)+ \frac{\sigma}{2P} \sum \limits_{j=i-P}^{i+P} (f(z_j^t)-f(z_i^t)) \right) + \nonumber \\
\alpha \left( f(z_i^{t+1})+ \frac{\sigma}{2P} \sum \limits_{j=i-P}^{i+P} . (f(z_j^{t+1})-f(z_i^{t+1})) \right) , \label{eq:9}
\end{eqnarray}
where $\alpha$ is a coefficient of anticipation.

The counterpart example to Eq.~\ref{eq:8} with `chimera' state is
\begin{eqnarray}
\frac{d \psi _k(t)}{dt} = \omega - \frac{1- \alpha}{2R} \sum \limits_{j=k-R}^{k+R} \sin (\psi _k(t)- \psi _j(t)+ \gamma ) + \nonumber \\
\frac{\alpha}{2R} \sum \limits_{j=k-R}^{k+R} \sin (\psi _k(t+ \tau )- \psi _j(t+ \tau )+ \gamma ). \label{eq:10}
\end{eqnarray}

It is accepted that the main source of existing of `chimera' states is the nonlocality in equations (see for example Eq.~\ref{eq:9}, Eq.~\ref{eq:10}). Remark that the nonlocality described in \cite{DanilenkoEtc} essentially expands the cases with presumable origing of `chimeras'. But the possibilities of `chimeras' in the systems with the anticipation are absolutely new. For example we should remark the possibilities of `multivalued chimeras', coexisting of different `chimeras' on different branches of the multivalued solutions, coexisting of `chimeras' and `smooth' behavior in the different branches of the solution.

\subsection{Presumable research of physical systems}
\label{subsec:45}

So far in this section we describe the possible properties of mathematical objects. Currently, the amount of experiments on the properties of the type of expression of the strict anticipation is very low (though they are made, especially in neuroscience and psychology). Therefore, we describe a hypothetical possibility in the physical realization of effects arising if we take the opportunity of physical realization of systems with strong anticipation property (there are several works in the favor of such opportunities, beginning with H.~Tetrode, R.~Feynman, D.~Dubois et al.).

\emph{Sources of the laws of probability}. In the case of physical realizability of strong anticipation, anticipation can not only be a model of probability laws, but also their physical mechanism.

\emph{The disruption of monitoring and instability stochastics in the experiments}. The experimental work with complex systems with stochastic processes in the experimental observation is sometimes observed failure monitoring, i.e. system suddenly jumps from one to another statistical laws (e.g., according to Victor Ivanenko, the author of systems with amplifiers). We can assume that the phenomenon could correspond to the jumps from one branch to another one of multi-valued solutions.

\emph{The appearance of an ensemble of systems in statistical physics, particularly in the Boltzmann scheme}. In statistical physics known Boltzmann distribution came from calculation with ensembles (configurations). However, in this scheme bands were purely speculative designs. Assuming strong anticipation one can assume that the configuration corresponding to the branches in the multivalued structure substantially parallel.

\emph{Problems of self-organization and synergy}. Classic problem of self-organization can also obtain new interpretations primarily because of possible ambiguity and change the language and the choice of appropriate functional unique implementation among many possible. In addition, an interesting problem is to assess the degree of disequilibrium in the future, making polysemy and branching.

\emph{Generalized quantum mechanics}, where the probabilities are set by anticipation. The usual quantum mechanics is the special case. In the classical scheme of quantum mechanics there is the equation for the probability density, which is complemented by the hypothesis of reduction (collapse) of the wave function of the measurements (the reduction of the system to a certain state). However the question still remains about the physical sense, the laws of probability and sources of quantum mechanics. Assuming the existence of a strong physical anticipation, one can consider the laws of probability generating mechanism with strong anticipation.

\emph{Everett's environment}. There exists Everett's interpretation of quantum mechanics, which assumes the existence of a plurality of parallel branching worlds. However, the mechanism of occurrence of such pattern has not yet been suggested. It makes sense to consider the possibility of generation of Everett's picture of the world through the mechanism of the type of strong anticipation. Generally, it is important to consider the environment with strong anticipation for which Everett pattern is a special case.

\emph{Cellular models of quantum fields}. It is known (G.~t~'Hooft, H.-T.~Elze, etc.) that the equations of classical quantum mechanics can be derived from models of cellular space on quantum scale (Planck scale) under limit transition to zero cell length. Consideration of strong anticipation in the original cellular models (recall the anticipation approach to Eq.~\ref{eq:9}, Eq.~\ref{eq:10} in Subsection~\ref{subsec:44}) can lead to differential equations for generalized quantum mechanics and just quantum gravity with anticipation. In such case the solutions of the processes of measurement correspond to choosing (or building) single-valued solution by measurement procedure.

\section{Possible ambiguity in the theory of computing machines and computer architecture}
\label{sec:5}

One of the most prospective new methodologies for modeling is the so called cellular automata approach. According to this method the models are built from the simple elements with local interaction with neighbors. The elements are distributed as the rule on simple geometry (lattices). Some details of approach to description of crowds and pedestrians will be described in next section. But here we recall the game `Life' by J.~Conway as the simplest example of CA. Recent investigations of cellular automata in physics and applied mathematics have shown that in spite of simple description of elements and rules for element dynamics they can represent any phenomena in nature -- self-organization, chaos, complex behavior of the whole system. The outstanding examples are modeling of hydrodynamic flows and many others. For example one of the CA applications is that simple (and frequently evident) rules that correspond to single pedestrian's behavior can lead to reliable modeling of crowd as complex object.

We describe new possibilities of CA in theory and applications including accounting anticipating (advanced equations). We propose both the examples of CA with anticipating and some applications.

We propose a list of some properties and manifestations of anticipatory property in different systems and processes: global sustainable development as strongly anticipative processes; regional sustainable development as weak anticipatory processes; origin of scenarios of evolution of complex social systems as the consequences of anticipation manifestation; self-referencing, reflexivity and mentality aspects in anticipation agents; medical manifestation of anticipation including schizophrenia; new consciousness models that are based on the anticipatory effects in the brain and artificial intelligence; quantum-mechanical, microphysics, gravitation and anticipatory analogies in the behavior of large complex systems.

Also the examples of anticipatory aspects in automata theory and computers have been considered. New scheme for probability investigation is proposed for considered systems with anticipation. Neural network models, cellular automata for crowd's movement, sport games, communication and social networks etc. are proposed.

\subsection{Presumable development of the theory of cellular automata}
\label{subsec:51}

In order to discuss the possible consequences of the new multivalued solutions in cellular automata we first of all will analyze the evolution of the computation theory and applications. After the initial period of automata theory development the next stage came in connection with the study of the game `Life' by J.~Conway. There were found many types of solutions of cellular automata, relation with the general theory of automata, the connection with the formal language theory and formal grammars, etc. The next stage began with the involvement of physics methods, primarily statistical physics methods in cellular automata studies. The beginning of application of physics methods in cellular automata in the 80s is associated with the names of S.~Wolfram, U.~Frisch, T.~Toffoli, B.~Chopard et al. for the simulation of hydrodynamics. Attractors were studied in parallel, for local and global maps, languages and machines related to cellular automata, the possibility of using the CA algorithms and programs, etc. Both deterministic and probabilistic CA and SA were constructed with a complicated structure (non-local, second-order (hierarchical), heterogeneous, with memory).

The next essentially new stage can be roughly related to the development of the theory of quantum computing including quantum cellular automata. The beginning of of this stage is usually connected with R.~Feynman's articles (1982, 1986) for quantum computation. The fundamentally important accomplishments are the formulation of the concept of a quantum machine, including CA, quantum Turing machines, quantum computation and logic applied to other quantum CA promising formulation local aspects CA in terms of local algebras of operators in the spirit of the results of G.~Haag.

Currently, all of the above areas continue to evolve (see the works by S.~Mura, J.~Crutchfield, T.~Toffoli, B.~Chopard and many others). Along the way are becoming increasingly popular and the use of CA \cite{Ilachinski,Wolfram}. Moreover, in view of hypotheses about the cellular structure of space-time at the micro-level (the Planck length, D.~Finkelshteyn's work (1973) and many others), it is sometimes declared that the universe is a giant cellular automaton with appropriate computational processes -- see the book by S.~Wolfram \cite{Wolfram} (2002).

Note that all the development of cellular automata occurred approximately as follows -- there were certain concepts (such as classical or quantum computing) that then were embedded into the concept of cellular automata. Of course, the proposed cellular automata with anticipation \cite{KrushinskyMakarenko} partially admit such an immersion into existing classical concepts. We call it the direct way of theory development. However, namely the solutions of the new CA models open the new paths for development of first of all theory that can be called the `inverse' or `reconstruction of concepts'.

\subsection{Problem of the reconstruction of concepts}
\label{subsec:52}

Again, now the development of cellular automata, both classical and quantum follows to `direct' way of development. That is, more often there were original concepts, then the CA were proposed, and then their properties and applications have been studied. Considered CA lead to the known equivalent representations (languages, automata, dynamical systems). All this is discussed in the framework of a predetermined structure of space-time. We can see that our research of CA with anticipation (and subsequent generalizations) has lead to new interesting results.

But there is a new source of problems, concepts, generalizations and interpretations. We call this way `reverse' in research or by `reconstruction' concepts. It comes from the fact that the primary, source, the basic element of the CA unit cell is given with its laws of evolution states. We assume that the existence of the enveloping concepts (meta--systems) and others with known properties were not originally supposed to, and they can be built, removed, identified only as properties of solutions of CA system. For example, for conventional CA were aware that such a machine, language, dynamic system, equivalence, etc. The introduction of anticipation in CA leads to ambiguity in the solutions decisions (we call these cellular automata with multivaluedness). To specify multiple meanings we will add the letter `M' to the corresponding object. Then the concepts of the theory of automata with multivaluedness (AM) should be introduced. Turing machine (TM) which allows ambiguity can be marked as (TMM), algorithms with multivaluedness -- (AlgM), formal languages with multivaluedness as (LM). Then it is necessary to revise the Turing thesis -- Church on computability by means of automata (TChM) Turing test for artificial intelligence (AIMtest).

In physics, this leads to the possibility of considering the multi-valued statistical mechanics (SMM); multivalued cosmology (KM), multi-valued quantum mechanics (CMM) and the multiple structures of matter, etc. Note that Everett's interpretation of quantum mechanics can be useful for such considerations. Conversely, multi-valued CA can help in understanding Everett's picture of the Universe. It seems that the new opportunities and understanding of the nature of probability can be proposed. More precisely, it can lead to new research on multiple-valued solutions in CA and evolutionary objects: multivalued chaos, cycles, bifurcations, Markov chains, attractors.

These are relatively obvious examples of `reconstruction' concepts. But no limits for generalizations exist. In fact, the structure of the `reverse' direction of research in CA can follow to generalized elements. Therefore, generalizations, first, can concern the state spaces of CA (or configurations). Also integer and real numbers can changes on other options (non-standard analysis, $p$-adic analysis; $\infty$-structure, logical calculus, topological objects (e.g. homology) distribution). Well, then state operators, secondary operators (operads), etc. can be proposed. To all this diversity more general rules -- generalized operators, multi-valued, discrete or continuous delay (and ahead) can be considered. There can be different definitions of neighborhoods (including implicit, fuzzy, time-dependent and state). Further generalizations can be viewed by using the CA as CA cells with elements of artificial intelligence. Thus, elements of the CA can be self-reflective elements with infinite recursion level. Each proposed architecture can induce construction of new structures, etc. We note only that the obvious way on this path suggests the language of categories and functors (including facilities for the study of the limit), which is becoming a commonplace in theoretical computer science and quantum field theory. We also point out that many of the issues discussed can also be applied to the model equations in the form of another type (e.g., differential).

Some features of cellular automata with strong anticipation are described in \cite{KrushinskyMakarenko}. We stress here only some interesting features. Firstly, we mention the application in the theory of automata and cellular automata. Cellular automata with strong anticipation can implement non-classical logic and can be non-Turing machines which are important for hypercomputation. In particular, possible worlds approach S.~Kripke and J.~Hintikka, as well as in line with the ideas of N.~Belnap can be implemented. The consideration of the implementation of computers in strong anticipative elements is promising. Also it is possible to consider the issues in the consciousness and effects on cognitive science \cite{Makarenko2012}.

\section{Interpretation of mental reactions and the problem of consciousness}
\label{sec:6}

Recently, studies of processes in the brain, proposing their concepts and models are attracting more and more attention. This list of achievements is impressive: a model of a single neuron, the theory of neural networks, neuroinformatics, cognitive science, brain--computer interface and many others.

However, it is still far from a final solution (and even from the existence of a common paradigm) because of the object complexity. One of the key problem is the explanation of the relationship of such entities as the brain --  mind -- consciousness.

In view of these uncertainties, it seems appropriate to consider different factors and phenomena, including those that have not yet been fully recognized. Naturally, there are a large number of factors. However, one of the most promising is the manifestation of the foresight properties (or more precisely of anticipation in the sense indicated below).

In this regard, the proposed paper describes some of the already established facts about anticipation in living systems, the properties of neural networks with anticipation and especially the possible consequences for the consideration of the problem of consciousness.

\subsection{Property of anticipation (foresight) in theory and nature}
\label{subsec:61}

There are many options to describe anticipation. Perhaps, for intuitive understanding the closest explanation is: `Anticipation (from the Latin. Anticipatio -- anticipate) -- presentation of the result of a process that occurs before its real achievements and serving as a means of feedback in the construction activities". This concept met, though without formalizing and measuring many times in the context of philosophy, economics, psychology, and medicine. More recently (probably in the last 30-40 years) anticipation emerged as experimental research and new theoretical concepts. In the neurosciences the most famous investigations were made by B.~Libet and numerous subsequent studies (the review can be found for example in \cite{Dubois2010} and many others).

On the theoretical level in the field of biology and modeling explicitly systems and models have been described anticipation R.~Rosen. Significant development and formalization of the anticipation concept was introduced by D.~Dubois.

One of the accepted concepts in classical neuroscience is to consider the brain processes in paradigm of neural network architecture for activity of the brain. Here we describe some presumable consequences of strong anticipation accounting in such models of brain activity.

\subsection{The possible role of strong anticipation in thinking}
\label{subsec:62}

The Subsection~\ref{subsec:35} briefly illustrates the results of the study of neural network models based on strong anticipation. The main new feature of this behavior compared to the other models is a multi-valued solution in the sense of simultaneous multiple virtual item states. Naturally, the simulation results indicate the need for further search analogues and manifestations of such effects in actual experiments. However, even the current results indicate a possible utility of the concept of powerful anticipation. So here we present some of the possible consequences of the adoption of the existence of effects such as strong anticipation.

Firstly let's discuss the problem of consciousness. Currently, there are a lot of concepts of consciousness such as psychological, biological, philosophical and physical (unfortunately, it is impossible to give a reference in such a short paper). We describe how the phenomenon of consciousness might look like in case of the strong anticipation accounting. The physical elements of the brain on different levels (from neurons to microtubules and Hammeroff cells) can take a lot of virtual values states. An illustration of how it looks like can be found in \cite{MakarenkoGoldengorinKrushinsky} by cellular automata -- the game `Life' in view of the strong anticipation. Then the elementary act of consciousness is the realization of one of the plurality of virtual states. Incidentally, it is strongly reminiscent of usual quantum mechanics complete with a measurement process (especially in Everett's interpretation). Acceptance of this interpretation can be more convenient for real applications results. This concept can be applied, for example, to the problem of schizophrenia. Then schizophrenia can be associated with a disorder of the mechanism of choice of the `ordinary' person of the plurality of virtual personalities. Also a virtual set of states can be correlated with the uncertainty in the ways of mental processes. Also anticipation B.~Libet experiments can correspond to a slow unconscious evaluation of virtual values of states, which is then accompanied by a conscious reaction and choosing the single one from many possibilities.

\section{Uncertainty and probability}
\label{sec:7}

History of the spread of probabilistic concepts has more than three hundred years of development, consisting of a stream of outstanding discoveries, mathematical and physical interpretation of the results. However, the main problem of the sources, meaning, nature and adequacy of the model laws of probability do not have the final solution yet. This is evident by a steady stream of publications on those issues. Although nowadays there are a lot of generalizing reviews and monographs, the latest studies describe new problems and approaches to major issues. Following the first modern research and review conditionally possible to allocate such directions:
\begin{itemize}
\item the classical approach with equally outcomes (from the 18th century);
\item frequency approach (primarily associated with the name of R.~von Mises);
\item the theory of possibilities (the propensity theory);
\item axioms of probability theory associated with the name of A.~Kolmogorov;
\item logical approach;
\item probabilistic aspects of decision theory and game approach, including the upper and lower probability;
\item quantum probability;
\item nonprobabilistic uncertainty (V.I.~Ivanenko);
\item probability and foundations of physics;
\item applied statistics and others.
\end{itemize}

Generally speaking, these areas can be arbitrarily structured in four blocks.
\begin{enumerate}
\item The axiomatic construction of the foundations of the theory.
\item Computation of probability characteristics of distributions et al., particularly in the prior centuries (for example, Markov chains).
\item Identification of new problems and systems in which it is possible to use probabilistic schemes for the calculation of uncertainties.
\item The physical interpretation of probability structures (for example, in statistical physics).
\end{enumerate}

Recently, in the blocks 3 and 4 a large number of new ideas (such as $p$-adic probability) initiated the concept of parastatistics problems of probability; the decision-making approaches in quantum mechanics; multiple non-deterministic chaos. As a rule, these new applications of probability schemes appear in connection with the consideration of new mathematical problems, describing various real processes.

In this paper we very shortly remark new possibilities of research on the basis of uncertainty properties of systems with strong anticipation (ahead). The ideas described here are mostly relevant to the block 3 above, that the study of models in which the manifestation of the properties that simulate probability is possible. Basic design centered around the possible emergence of ambiguities in the evolution of solutions in problems with strong anticipation and arising in this uncertainty by multi-valuedness of solutions and emerging of many presumable pathes for given values of solutions \cite{Makarenko2013}. This can lead to the analog of classical `frequencies of realizations' in counting the probability. A general scheme of such structure and specific examples, particularly cellular automata and artificial neural network, are given. We also discuss the hypothesis of a possible physical realization of such systems and their relationship with the occurrence of the actual laws of probability.

\section*{Conclusions}
\label{sec:8}

Thus in given paper we propose to consider the problem of complex behavior for new class of objects namely for the chains and networks with strong anticipation. The presumable multivaluedness of solutions leads to new interesting properties in the frame of already existing concepts. But new features can also appear (for example non-heterogeneous multivalued synchronization) that are very promising for further investigation. We described only the first results of investigations and only some new presumable forms of research problems. But evident mathematical novelty of proposed problems and presumable great importance in applications (for example for social systems; computation and signal processing theory; consciousness investigations etc.) leads to the need of further development of investigations.

Thus, this paper describes examples of the introduction of anticipation in terms of cellular automata, neural networks, differential and discrete equations. One of the important results is the possibility of the emergence of multi-valued solutions. An immediate consequence is the possibility of posing problems of multivalued machines, languages, chaos, etc. Moreover, the results on multivalued solutions (including multivalued solutions by strong anticipation) described in the paper can lead to hypothesis for discussion that world can have the multivalued nature but we being the observers can usually see only the single-valued picture of the world.

\noindent {\it Acknowledgement.}
The author would like to give the appreciation for D.~Krusinsky, Yu.~Yatsuk, V.~Biliuga, A.~Stashenko and S.~Lazarenko for computer realizations of models and V. Statkevich for assistance in preparing the text of the paper.

\end{document}